\pdfoutput=1
\documentclass{article}

\usepackage[utf8]{inputenc} 
\usepackage{hyperref}       
\usepackage{url}            
\usepackage{amsfonts}       
\usepackage{nicefrac}       
\usepackage{microtype}      

\usepackage[a4paper]{geometry}
\usepackage{ctable}
\usepackage{graphicx}
\usepackage{subcaption}
\usepackage{amsmath}
\usepackage{pythonhighlight}

\begin{document}

\title{PyIT2FLS: A New Python Toolkit for Interval Type 2 Fuzzy Logic Systems}

\author{
  Arslan A. Haghrah \\
  Faculty of Electrical and Computer Engineering\\
  University of Tabriz\\
  Tabriz, Iran \\
  \texttt{haghrah@tabrizu.ac.ir} \\
\and
  Sehraneh Ghaemi \\
  Faculty of Electrical and Computer Engineering\\
 University of Tabriz\\
  Tabriz, Iran \\
  \texttt{ghaemi@tabrizu.ac.ir} \\
}
\maketitle

\begin{abstract}
Fuzzy logic is an accepted and well-developed approach for constructing verbal models. Fuzzy based methods are getting more popular, while the engineers deal with more daily life tasks. This paper presents a new Python toolkit for Interval Type 2 Fuzzy Logic Systems (IT2FLS).  Developing software tools is an important issue for facilitating the practical use of theoretical results. There are limited tools for implementing IT2FLSs in Python. The developed PyIT2FLS is providing a set of tools for fast and easy modeling of fuzzy systems. This paper includes a brief description of how developed toolkit can be used. Also, three examples are given showing the usage of the developed toolkit for simulating IT2FLSs. First, a simple rule-based system is developed and it's codes are presented in the paper. The second example is the prediction of the Mackey-Glass chaotic time series using IT2FLS. In this example, the Particle Swarm Optimization (PSO) algorithm is used for determining system parameters while minimizing the mean square error. In the last example, an IT2FPID is designed and used for controlling a linear time-delay system. The code for the examples are available on toolkit's GitHub page: \url{https://github.com/Haghrah/PyIT2FLS}. The simulations and their results confirm the ability of the developed toolkit to be used in a wide range of the applications.
\end{abstract}


\section{Introduction}
While the fuzzy logic had not attracted many researchers in the first years of its presentation, the usage of the fuzzy approaches in science and engineering applications has raised over time. By increasing the number of successful applications of this approach on real-world problems, research on analyzing and design of fuzzy systems has flourished \cite{tanaka2004fuzzy}. The key idea of the fuzzy systems is approximate reasoning, which is based on IF-THEN rules. A fuzzy IF-THEN rule is defined as a statement built of words, expressed by membership functions \cite{wang1997course}. Measuring the membership grade using crisp numbers yields type 1 fuzzy sets. But when the membership grade is also uncertain and is represented using fuzzy membership functions, the concept of type 2 and higher types of fuzzy sets are introduced \cite{zadeh1975concept}. It must be noticed that, by increasing the type of a fuzzy set, the computational burden significantly grows. Therefore researchers have introduced the concept of interval type 2 fuzzy sets for decreasing the computational load, in which the secondary membership value is always considered to be equal to one \cite{mendel2014introduction}.\\
Type-2 fuzzy systems are applied widely to solve engineering and industrial problems, which among them control of buck and boost DC-DC converters \cite{atacak2012type}, coiler entry temperature prediction \cite{mendez2010modelling}, levelling the cable-driven parallel mechanism \cite{cheng2010inverse}, control of induction motor \cite{barkati2008application}, and digital image filtering \cite{singh2018adaptive} can be mentioned. Similar applications of type 2 fuzzy systems are advancing rapidly. However, compared with type 1 fuzzy systems, type 2 fuzzy systems have been used very limitedly \cite{wagner2013juzzy}. Public access and use of an approach, derived from theoretical results, is mainly affected by the availability of software implementations. \\
There are some toolkits developed for utilizing type 2 fuzzy sets and systems in science and engineering applications. In \cite{castillo2013computational}, a new MATLAB toolbox for interval type 2 fuzzy logic systems has been introduced. In addition to basic features for designing IT2FLSs, this toolbox includes a graphical user interface for construction, edition, and observation of the fuzzy systems. Also, the mentioned paper studies both educational and research sides of developing the toolbox. The results of the experiments, in addition to the effectiveness of the toolbox, indicate the technical improvement of the students who have used the tool. Another toolbox which is designed to be used with MATLAB, like the previous one, has been presented in \cite{Taskin2015it2fls}. In this toolbox, in addition to the graphic user interface for designing and editing fuzzy systems, a Simulink model is also presented which can be mentioned as its advantage. As an example of using the toolbox, in this paper, an IT2-Fuzzy PID controller is designed for controlling a system in the presence of time delay. Another Java-based fuzzy logic toolkit, named Juzzy, is presented in \cite{wagner2013juzzy}. Juzzy supports type 1, interval type 2 and general type 2 fuzzy logic systems. The aim of the developers of Juzzy toolkit is facilitating the application of fuzzy logic systems in real-world problems, along with improving accessibility to the type 2 fuzzy logic systems for academic purposes. Recently a new Python-based toolkit for automatically generating and analyzing data-driven fuzzy sets has been developed and presented in \cite{McCulloch2017}. The introduced toolkit, fuzzycreator, provides many routines for creating conventional and non-conventional type 1, interval type 2 and general type 2 fuzzy sets from data. It must be noticed that the developed toolbox focuses on fuzzy sets rather than fuzzy systems.

Due to the increasing popularity of the Python programming language and its frequent use for scientific and engineering applications, the need for a tool for providing IT2FLSs support has been felt. To respond to this need, the PyIT2FLS toolkit is developed and presented in this paper. This toolkit provides many useful functions for creating and processing fuzzy sets and systems. There are two core classes, which the features of the toolbox are available through them, \textit{IT2FS} and \textit{IT2FLS}. The first one for modeling the fuzzy sets and the latter for the systems. It must be noticed that the ease of use is considered as the main criterion in the development of the toolkit. In the rest of the paper when a keyword refers to a function, variable or class it is written in italic form.

The rest of the paper is organized as follows. The second section consists of a brief discussion on the toolkit and how to use it. In Section 3, three examples of using the PyIT2FLS are provided. Finally, Section 4 presents some conclusions.

\section{PyIT2FLS}
This section presents how to use the new developed Python toolkit, PyIT2FLS, for modeling and simulation of interval type 2 fuzzy systems. PyIT2FLS is developed to respond to the lack of an integrated tool for facilitating the use of IT2FLSs. This toolkit, which its development will not stop in this stage, will allow the users to apply IT2FL based methods for solving issues encountered in science and engineering problems. PyIT2FLS is based on well-known numeric library NumPy \cite{oliphant2006guide} and plotting library Matplotlib \cite{hunter2007matplotlib}. It consists of two main classes, \textit{IT2FS} for interval type 2 fuzzy sets and \textit{IT2FLS} for interval type 2 fuzzy logic systems, and many utility functions concerning the interval type 2 fuzzy logic. As a starting point, we introduce some of the utility functions defined in the toolkit, which are used in creating interval type 2 fuzzy sets. These are general membership functions, which would be used as the lower membership function (LMF) and the upper membership function (UMF) for \textit{IT2FS}s. Table \ref{mfTable} represents these functions with a brief description. All these functions have two inputs, vector-like input $x$ indicating points from the domain which the function must be evaluated, and $params$ indicating the function's parameters. It should be noted that the latter parameter must be a list, and for each function its items are described in its corresponding docstring.
\begin{table}
 \caption{List of functions which can be used as LMFs and UMFs.}
  \centering
  \begin{tabular}{ll}
    \hline
    Name			& Description \\
    \hline
    zero\_mf 		& All zero membership function \\
    singleton\_mf	& Singleton membership function \\
    const\_mf		& Constant membership function  \\
    tri\_mf			& Triangular membership function  \\
    trapezoid\_mf	& Trapezoidal membership function  \\
    gaussian\_mf		& Gaussian membership function  \\
    gauss\_uncert\_mean\_umf	& UMF of Gaussian membership function with uncertain mean value \\
    gauss\_uncert\_mean\_lmf	& LMF of Gaussian membership function with uncertain mean value \\
    gauss\_uncert\_std\_umf		& UMF of Gaussian membership function with uncertain std. value \\
    gauss\_uncert\_std\_lmf		& LMF of Gaussian membership function with uncertain std. value \\
    \hline
  \end{tabular}
  \label{mfTable}
\end{table}

\subsection{Interval type 2 fuzzy sets}

The \textit{IT2FS} class defines interval type 2 fuzzy sets by specifying the domain, UMF, parameters of UMF function, LMF, and parameters of LMF function. The first example in appendix shows how to define an \textit{IT2FS} using aforementioned membership functions. The only constraint for using these functions is to choose the function parameters in such a way that for all points in the universe of discourse, $UMF(x) > LMF(x)$ be satisfied.

\begin{figure}
  \centering
  \includegraphics[width=5cm]{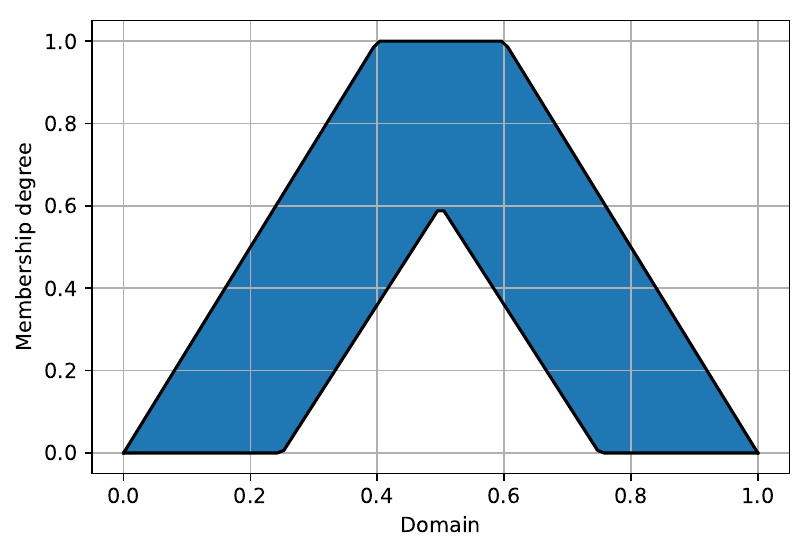}
  \caption{An IT2FS with trapezoidal UMF, and triangular LMF.}
  \label{mySet}
\end{figure}

The set defined using the code is depicted in Figure. \ref{mySet}. There are two conventional types of IT2FSs which can be defined easily using two other functions available in PyIT2FLS. In order to define Gaussian IT2FS with uncertain mean value and uncertain standard deviation value, users can call $IT2FS\_Gaussian\_UncertMean$ and $IT2FS\_Gaussian\_UncertStd$ functions, respectively. The parameters of the both functions are $domain$ and $params$. For the first function, the second parameter consists of the mean value center, mean value spread, and standard deviation and for the second function, the second parameter consists of the mean value, standard deviation center, and standard deviation spread (for both as lists). Each \textit{IT2FS} in PyIT2FLS contains some parameters and functions which are indicated in Table \ref{IT2FS_params} and Table \ref{IT2FS_funcs}, respectively. More detailed descriptions and instruction for each of them are available in docstrings.
\begin{table}
 \caption{List of parameters defined in IT2FS class.}
  \centering
  \begin{tabular}{ll}
    \hline
    Parameter		& Description\\
    \hline
    domain		& Defines the universe of discourse\\
    umf 		& Defines upper membership function\\
    umf\_params	& Defines upper membership function parameters\\
    lmf 		& Defines lower membership function\\
    lmf\_params	& Defines lower membership function parameters\\
    upper 		& Upper membership function values in the discrete universe of discourse\\
    lower		& Lower membership function values in the discrete universe of discourse\\
    \hline
  \end{tabular}
  \label{IT2FS_params}
\end{table}
\begin{table}[htp]
 \caption{List of functions defined in IT2FS class.}
  \centering
  \begin{tabular}{ll}
    \hline
    Function		& Description \\
    \hline
    copy			& Returns a copy of the IT2FS \\
    plot 			& Plots the IT2FS in the universe of discourse \\
    \hline
  \end{tabular}
  \label{IT2FS_funcs}
\end{table}
The next function to be introduced is $IT2FS\_plot$ which plots a number of \textit{IT2FS}s together in the same figure. The main inputs of this function are a number of \textit{IT2FS}s, but title and legends can also be assigned. Furthermore if the user wishes to save the plotted figure in $pdf$ format, he/she can set the \textit{filename} input variable. More detailed description and examples are available in docstrings. 
\subsection{T-norms, s-norms, join and meet}
\begin{figure}
\begin{subfigure}{0.33\linewidth}
\centering
\includegraphics[width=4.5cm]{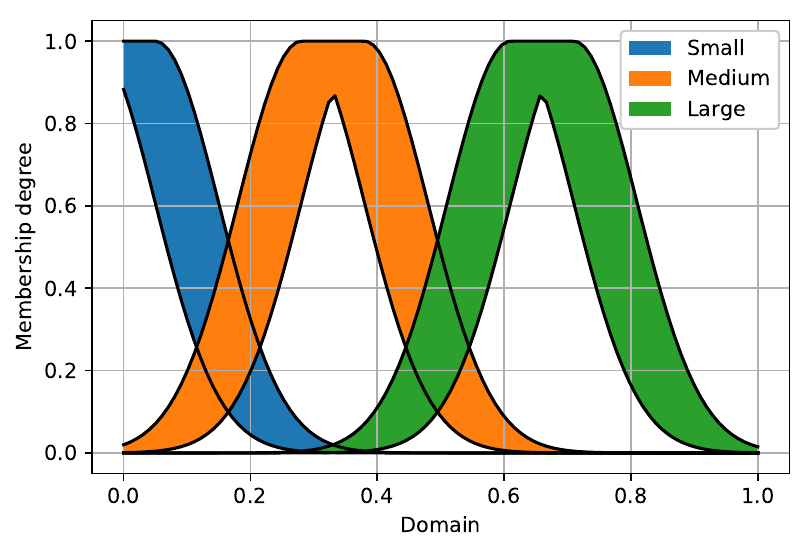}
\caption{The IT2FSs used in example.}
\label{fig:meetjoin_sets}
\end{subfigure}
\begin{subfigure}{0.33\linewidth}
\centering
\includegraphics[width=4.5cm]{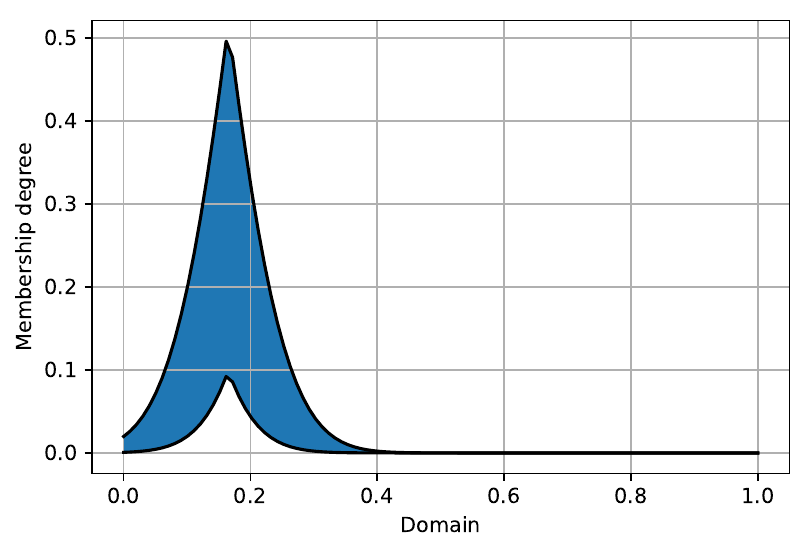}
\caption{Meet of Small and Medium sets.}
\label{fig:meet}
\end{subfigure}%
\begin{subfigure}{0.33\linewidth}
\centering
\includegraphics[width=4.5cm]{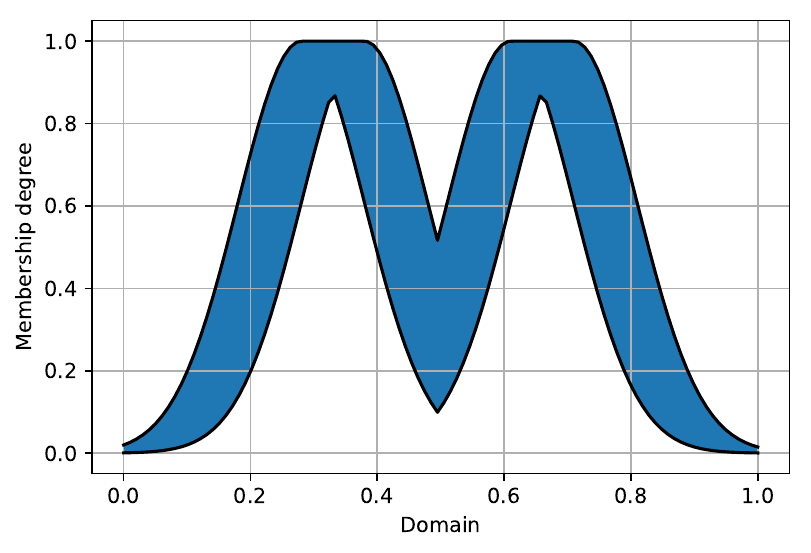}
\caption{Join of Medium and Large sets.}
\label{fig:join}
\end{subfigure}\\[1ex]
\caption{Sample outputs for \textit{IT2FS\_plot}, meet and join functions.}
\label{fig:meetjoin}
\end{figure}

As basic operators in fuzzy calculations, t-norms and s-norms have special importance. PyIT2FLS by default contains two t-norm functions, $min\_t\_norm$ and $product\_t\_norm$ for minimum and product t-norms respectively, and a s-norm, $max\_s\_norm$ for maximum s-norm. However, new t-norm and s-norm functions can be defined by users. In the standard form, these functions must have two array-like inputs for calculating t-norm or s-norm, element-wise on them. The next two main operators in type 2 fuzzy logic are meet and join operators which are defined by two functions with the same names, \textit{meet} and \textit{join}. The inputs of these operators are domain, first and second IT2FS, and norm operator. For the \textit{meet}, norm operator must be a t-norm and for the \textit{join}, norm operator must be a s-norm. The second code in the appendix shows how to use meet and join operators and plotting multiple sets together. The results are demonstrated in Figure \ref{fig:meetjoin}.

\subsection{Interval type 2 fuzzy logic systems}
The most important part of the developed toolkit, the \textit{IT2FLS} class, is prepared for modeling the interval type 2 fuzzy logic systems. Each \textit{IT2FLS} has three parameters, \textit{inputs}, \textit{outputs}, and \textit{rules}. There are five functions defined in \textit{IT2FLS} class for utilizing it which are presented in Table \ref{funcs_IT2FLS}. No construction parameters are needed when an \textit{IT2FLS} is created. The input parameters of the functions for adding inputs and outputs are only the names. In order to define the rule base of the fuzzy system $add\_rule$ function is used. This function accepts two lists as input which the first one indicates antecedent part of the rule and the second one indicates conclusion part of the rule. Each list consists of tuples indicating the attachment of input variables to corresponding \textit{IT2FS}s. Assume that a fuzzy type 2 IF-THEN rule be as below:
\begin{equation}
\begin{aligned}
\text{IF}\quad &x_{1}\ is\ \tilde{A}_{1}\quad \text{AND}\quad x_{2}\ is\ \tilde{A}_{2}\quad \text{AND}\quad ... \quad \text{AND}\quad x_{n}\ is\ \tilde{A}_{n}\quad \text{THEN}\\
&y_{1}\ is\ \tilde{B}_{1},\quad y_{2}\ is\ \tilde{B}_{2},\quad ... \quad y_{m}\ is\ \tilde{B}_{m}
\end{aligned}
\end{equation}
So this rule can be added to an \textit{IT2FLS} named \textit{myIT2FLS} as pseudo-code below:
\begin{equation}
\begin{aligned}
myIT2FLS.add\_rule(&[("x1",\ A1),\ ("x2",\ A2),\ ...,\ ("xn",\ An)], \\
                   &[("y1",\ B1),\ ("y2",\ B2),\ ...,\ ("ym",\ Bm)]) \nonumber
\end{aligned}
\end{equation}
In order to evaluate an \textit{IT2FLS} for arbitrary inputs the \textit{evaluate} function must be used. The inputs of this function are values of the system inputs as a dict, t-norm, s-norm, domain, and the type reduction algorithm and method. The default output of the \textit{evaluate} function is depended on the type reduction method and the algorithm specified by the user and is described in details in docstrings. 

In most of cases the output variable of this function is of dict data type. In the output dict, tuples are corresponded with the output variable names which are added by \textit{add\_output\_variable} function in initial definition of the system. These tuples represent type reduced outputs of the fuzzy system. In order to obtain crisp output of the system, each tuple can be given as input of the \textit{crisp} function. Also, in order to plot the type reduced output the \textit{TR\_plot} function cab ne used. The third code in the appendix section shows how to plot the type reduced outputs and calculate the crisp outputs for the result of \textit{evaluate} function.

\begin{table}
 \caption{List of functions defined in IT2FLS class.}
  \centering
  \begin{tabular}{ll}
    \hline
    Function		& Description \\
    \hline
    add\_input\_variable	& Adds a new input variable to IT2FLS \\
    add\_output\_variable		& Adds a new output variable to IT2FLS \\
    add\_rule			& Adds a new rule to IT2FLS's rule base \\
    evaluate			& Evaluates the IT2FLS with specific inputs \\
    copy			& Returns a copy of the defined IT2FLS \\
    \hline
  \end{tabular}
  \label{funcs_IT2FLS}
\end{table}

\subsection{Type reduction}
A very important concept in type 2 fuzzy systems is type reduction algorithms and methods. Nine type reduction algorithms are provided in PyIT2FLS which are KM \cite{karnik1999type}, EKM, WEKM, TWEKM \cite{duran2008improved, wu2008enhanced}, EIASC\cite{wu2011comparison}, WM \cite{wu2002uncertainty}, BMM \cite{begian2008stability}, LBMM \cite{li2010stability}, and NT \cite{nie2008towards}. These type reduction algorithms can be used alongside the methods Centroid, Center of sets, Center of sum, Height, and Modified height. The functions corresponding with these algorithms and methods and their brief description are presented in Tables \ref{TypeRe_algs} and Table \ref{TypeRe_methods}, respectively. These functions and methods are not called directly by the user most of the times, but their name is passed to the \textit{evaluate} function of \textit{IT2FLS} class as parameters, \textit{method} and \textit{algorithm}. The aforementioned tables also contain the names which the algorithms and methods are passed to the \textit{evaluate} function. Docstrings are also available for a more detailed explanation of the instructions.
\begin{table}
 \caption{List of type reduction algorithms for \textit{evaluate} function of \textit{IT2FLS} class.}
  \centering
  \begin{tabular}{lll}
    \hline
    Algorithm		& Description & Name\\
    \hline
    KM		& Karnik-Mendel algorithm  &KM \\
    EKM		& Enhanced KM algorithm  &EKM \\
    WEKM	& Weighted EKM algorithm  &WEKM \\
    TWEKM	& Trapezoidal WEKM algorithm   &TWEKM \\
    EIASC	& Enhanced IASC algorithm  &EIASC \\
    WM		& Wu-Mendel algorithm  &WM \\
    BMM		& Begian-Melek-Mendel algorithm  &BMM \\
    LBMM	& BMM method edited by Li et al.  &LBMM \\
    NT		& Nie-Tan algorithm   &NT \\
    \hline
  \end{tabular}
  \label{TypeRe_algs}
\end{table}

\begin{table}
 \caption{List of type reduction methods for \textit{evaluate} function of \textit{IT2FLS} class.}
  \centering
  \begin{tabular}{lll}
    \hline
    Method				& Description 								& Name\\
    \hline
    Centroid			& Centroid type reduction method  			& Centroid\\
    Center of sets		& Centr of sets type reduction method 		& CoSet\\
    Center of sum		& Center of sums type reduction method 		& CoSum\\
    Height				& Height type reduction method 				& Height\\
    Modified height		& Modified height type reduction method 	& ModiHe\\
    \hline
  \end{tabular}
  \label{TypeRe_methods}
\end{table}

\section{Examples}
In this section three examples of using proposed PyIT2FLS are under study. First a simple example for using the toolkit to evaluate fuzzy IF-THEN rules as a fuzzy logic system is given. After that an example of using the toolkit for prediction of chaotic time series is given, in which the parameters of fuzzy system are determined using PSO algorithm to have minimum mean square error. And finally, the toolkit is used for modeling an interval type 2 fuzzy PID (IT2FPID) controller, and using it to control a time-delay linear system.
\subsection{A simple example for using PyIT2FLS}
Assume an IT2FLS with two inputs and two outputs. Let the universe of discourse in this example be the interval $[0,\ 1]$. Assume that three IT2FSs are defined in the universe of discourse which are Gaussian IT2FS with uncertain standard deviation values. Also let the rule base of the IT2FLS be as below:
\begin{equation}
\begin{aligned}
&\text{IF}\ \  x_{1}\ \  is\ \  \text{Small}\quad \ \ AND\ \  x_{2}\ \  is\ \  \text{Small} \ \ \ \quad \text{THEN}\quad y_{1}\ \  is\ \  \text{Small}\ \ \quad AND\ \  y_{2}\ \  is\ \  \text{Large} \\
&\text{IF}\ \  x_{1}\ \  is\ \  \text{Medium}\ \  AND\ \  x_{2}\ \  is\ \ \text{Medium} \ \ \text{THEN}\ \ \ y_{1}\ \  is\ \ \text{Medium}\ \  AND\ \  y_{2}\ \  is\ \  \text{Small}\\
&\text{IF}\ \  x_{1}\ \  is\ \ \text{Large}\quad \ \ AND\ \  x_{2}\ \  is\ \  \text{Large}\ \ \ \quad \text{THEN}\quad y_{1}\ \  is\ \ \text{Large}\ \ \quad AND\ \  y_{2}\ \  is\ \  \text{Small} \nonumber
\end{aligned}
\end{equation}
With these assumptions, the codes for modeling this IT2FLS would be as provided in fourth subsection of the appendix. The discrete domain is defined as 100 equal-distance points in the interval $[0, 1]$. The sets defining the universe of discourse are constructed using the function \textit{IT2FS\_Gaussian\_UncertStd} and are plotted. The inputs $x_1$ and $x_2$, and the outputs $y_1$ and $y_2$ are added to the system. Based on defined \textit{IT2FS}s and input and output variables, fuzzy IF-THEN rules are defined. Finally the output is evaluated and represented for $x_1,\ x_2=0.9$.

\begin{figure}
\begin{subfigure}{0.23\linewidth}
\centering
\includegraphics[width=3.5cm]{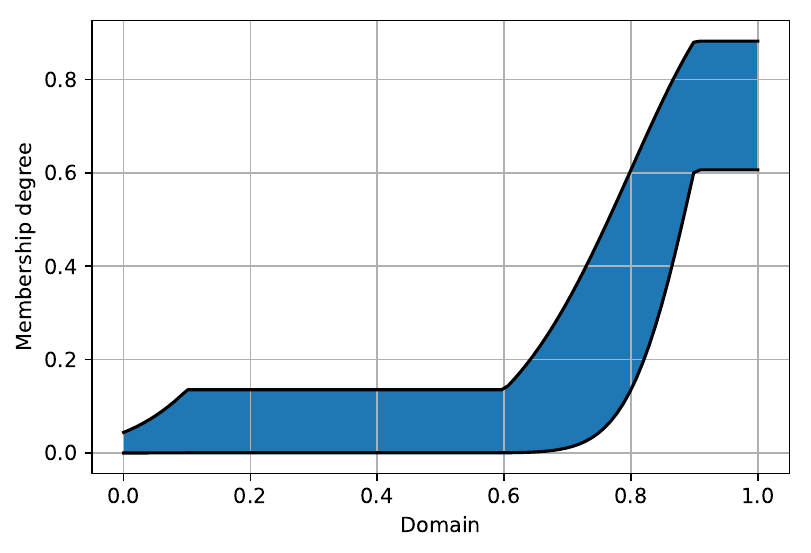}
\caption{$y_1$}
\label{fig:ex_out1}
\end{subfigure}%
\begin{subfigure}{0.23\linewidth}
\centering
\includegraphics[width=3.5cm]{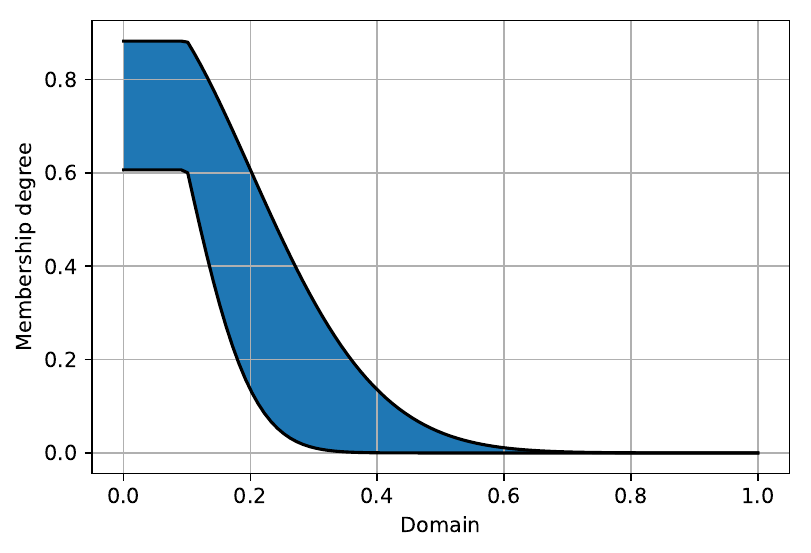}
\caption{$y_2$}
\label{fig:ex_out2}
\end{subfigure}
\begin{subfigure}{0.23\linewidth}
\centering
\includegraphics[width=3.5cm]{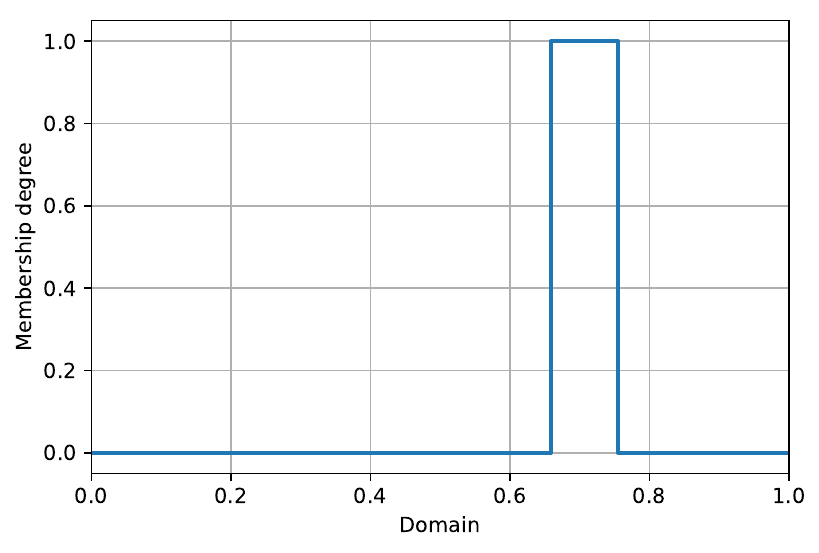}
\caption{Type reduced $y_1$}
\label{fig:tr_ex_out1}
\end{subfigure}%
\begin{subfigure}{0.23\linewidth}
\centering
\includegraphics[width=3.5cm]{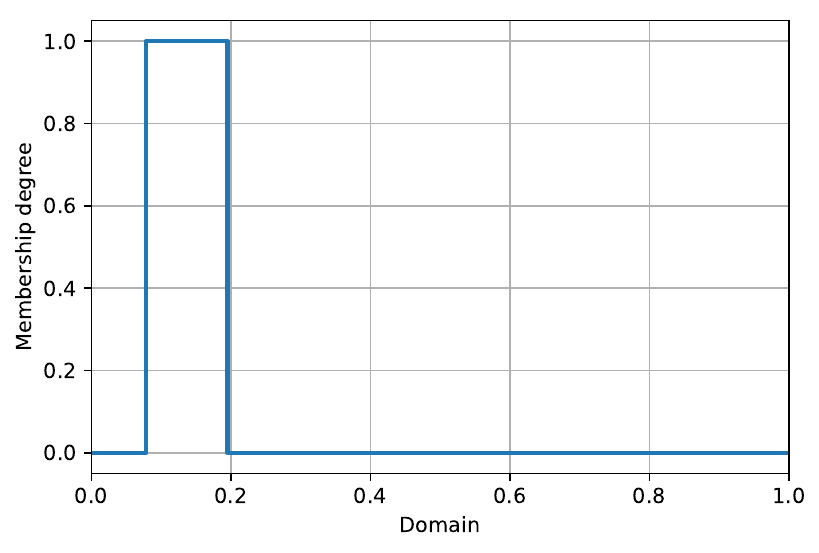}
\caption{Type reduced $y_2$}
\label{fig:tr_ex_out2}
\end{subfigure}\\[1ex]
\caption{Outputs of the provided example.}
\label{fig:ex_outs}
\end{figure}

The outputs generated by running this code are demonstrated in Figure. \ref{fig:ex_outs}. IT2FS output for first and second outputs of the system are illustrated in Figures \ref{fig:ex_out1} and \ref{fig:ex_out2}, respectively. Also type reduced outputs are shown in Figures \ref{fig:tr_ex_out1} and \ref{fig:tr_ex_out2}.


\subsection{Prediction of Mackey-Glass Chaotic Time Series}
\begin{figure}
  \centering
  \includegraphics[width=5cm]{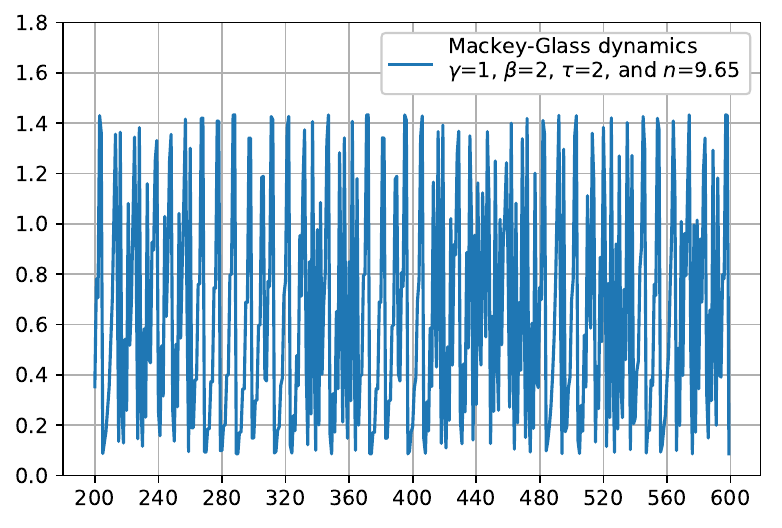}
  \caption{Response of the Mackey-Glass nonlinear time delay differential equation for exact parameters.}
  \label{mackeyglass1}
\end{figure}

In this section, the results achieved by using the developed PyIT2FLS in a practical application, prediction of Mackey-Glass chaotic time series, is shown. The Mackey-Glass nonlinear time delay differential equation is defined as:
\begin{equation}
\frac{dx}{dt} = \beta \frac{x_{\tau}}{1 + x_{\tau}^{n}} - \gamma x \qquad \gamma, \beta, n>0
\end{equation}
in which the parameters $\beta$, $\gamma$, $\tau$ and $n$ are real numbers. In addition, $x_{\tau}$ shows the delayed value of variable $x$ at time $t-\tau$ \cite{glass2010mackey}. The response of the Mackey-Glass dynamic equation is demonstrated in Figure. \ref{mackeyglass1} for parameters $\beta=2$, $\gamma=1$, $\tau=2$, and $n=9.65$ and random initial conditions.

The Mackey-Glass Equation alongside the logistic map, Lorenz equations, and Chen systems is known as a paragon of chaotic systems \cite{roussel2018mackey}. Due to the chaotic behavior of the Mackey-Glass time series, it has been the subject of many scientific studies. These studies mainly concern the prediction of time series \cite{zhongda2017prediction, heydari2016chaotic} and the synchronization of chaotic systems \cite{shahverdiev2006chaos, shi2017synchronization}. 

\begin{figure}
\begin{subfigure}{0.48\linewidth}
\centering
\includegraphics[width=5cm]{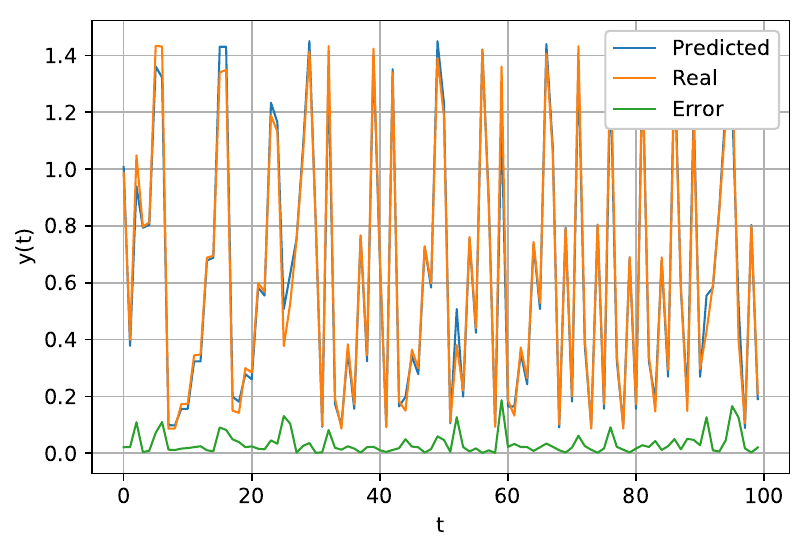}
\caption{Prediction output achieved using the PyIT2FLS.}
\label{mg_out}
\end{subfigure}%
\begin{subfigure}{0.48\linewidth}
\centering
\includegraphics[width=5cm]{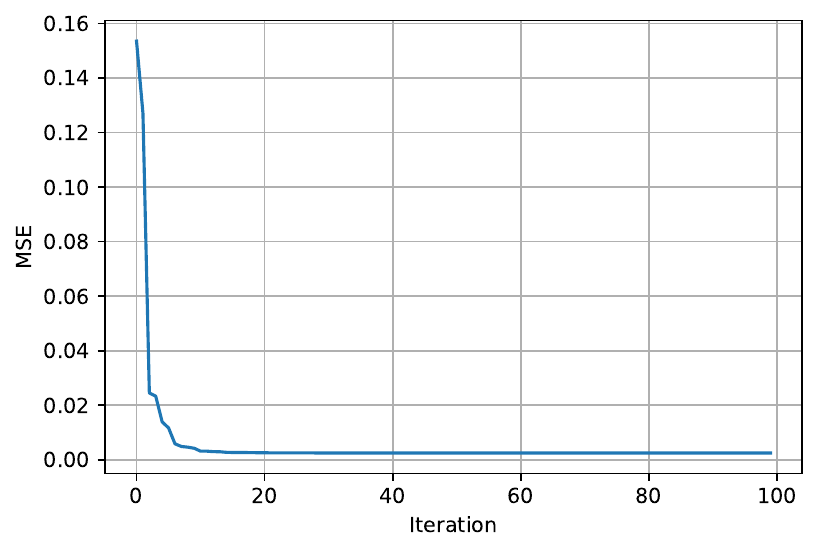}
\caption{Convergence of the PSO algorithm.}
\label{convergence}
\end{subfigure}\\[1ex]
\caption{Outputs achieved for the predictiin of Mackey-Glass time series using PyIT2FLS.}
\label{mg_simuresult}
\end{figure}

Let $mg$ indicate the time series, in order to predict the Mackey-Glass time series using IT2FLS, three inputs are considered which are consecutive samples $mg[t-2]$, $mg[t-1]$, and $mg[t]$. Assume that inputs are named as A, B, and C. Each input in its universe of discourse is expressed by three IT2FSs, which can be shown by indices 1 to 3 (as an example for the input A there will be three sets $A_1$, $A_2$, and $A_3$). The output of system, named O, is $mg[t+1]$ and is expressed using three IT2FSs, $O_1$, $O_2$, and $O_3$. All the fuzzy sets are Gaussian IT2FS with uncertain standard deviation values. The rule base of the system is demonstrated in Table \ref{RuleBase}. All the parameters of the fuzzy sets are selected by PSO algorithm to have minimum Mean Square Error (MSE) in prediction. In this step, 100 samples are used for optimizing the selected parameters. Also the prediction is repeated on an other 100 samples in order to guarantee the performance of the IT2FLS. The result of prediction, real value of time series and absolute error are illustrated in Figure \ref{mg_out}. Further more the convergence diagram of the PSO algorithm is demonstrated in Figure \ref{convergence}. The codes used for simulating this example are openly accessible over the toolkit's GitHub page.

\begin{table}
 \caption{Fuzzy If-Then rule base of IT2FLS.}
  \centering
  \begin{tabular}{ccccc}
    \hline
    Rule &\multicolumn{3}{c}{Antecedent} &Conclusion\\
     &A &B &C &O\\
    \hline
    Rule 1 & A1 & B1 & C1 & O1\\
    Rule 2 & A2 & B2 & C2 & O2\\
    Rule 3 & A3 & B3 & C3 & O3\\
    \hline
  \end{tabular}
  \label{RuleBase}
\end{table}

\subsection{Interval Type 2 Fuzzy PID Controller for Time Delay Linear System}
\begin{figure}
  \centering
  \includegraphics[width=16cm]{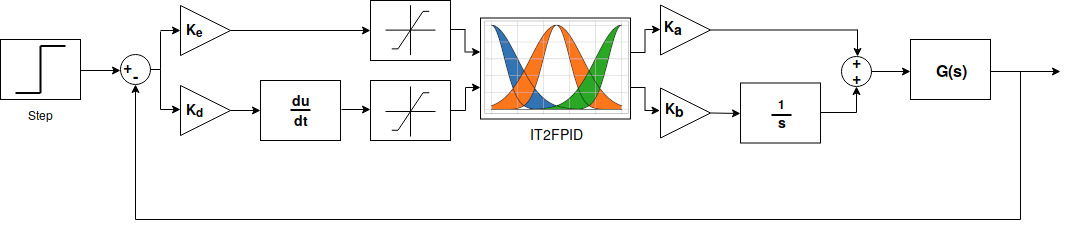}
  \caption{Block diagram of the system under study with IT2FPID controller.}
  \label{blockdiag}
\end{figure}

\begin{figure}
\begin{subfigure}{0.5\linewidth}
\centering
\includegraphics[width=5cm]{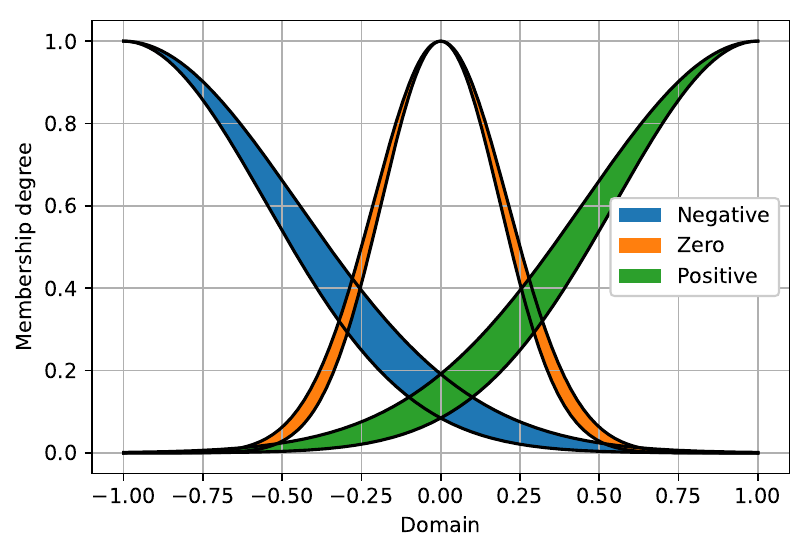}
\caption{Sets defined for the inputs of IT2FPID}
\label{i_IT2FPID_Sets}
\end{subfigure}%
\begin{subfigure}{0.5\linewidth}
\centering
\includegraphics[width=5cm]{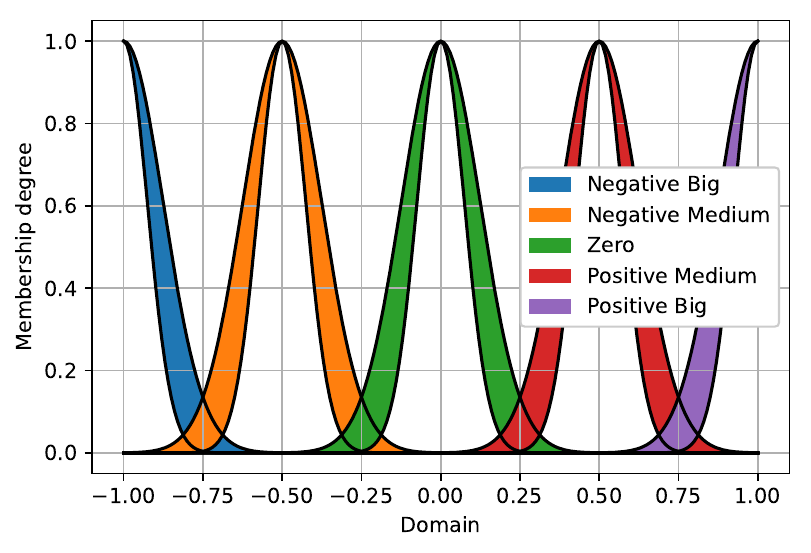}
\caption{Sets defined for the output of IT2FPID}
\label{o_IT2FPID_Sets}
\end{subfigure}\\[1ex]
\caption{Input sets and output sets defined for PyIT2FLS.}
\label{io_IT2FPID_Sets}
\end{figure}

\begin{figure}
\begin{subfigure}{0.32\linewidth}
\centering
\includegraphics[width=5cm]{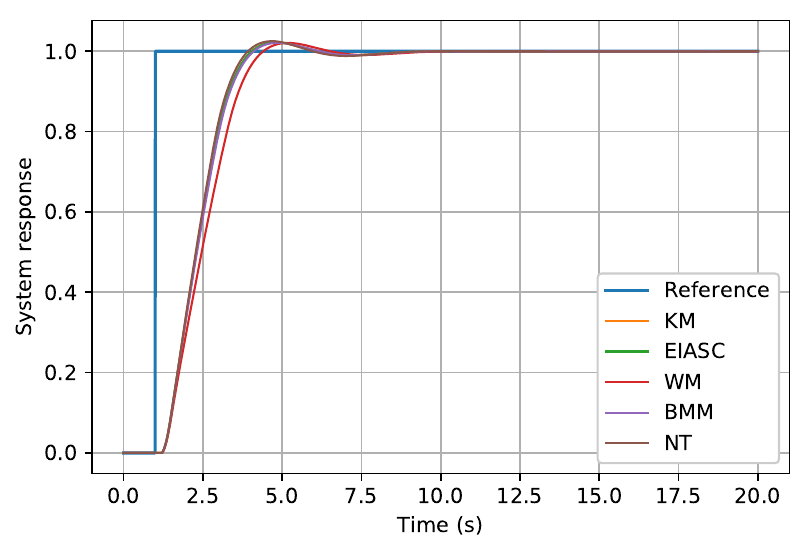}
\caption{Main system}
\label{FPID_Plant1}
\end{subfigure}%
\begin{subfigure}{0.32\linewidth}
\centering
\includegraphics[width=5cm]{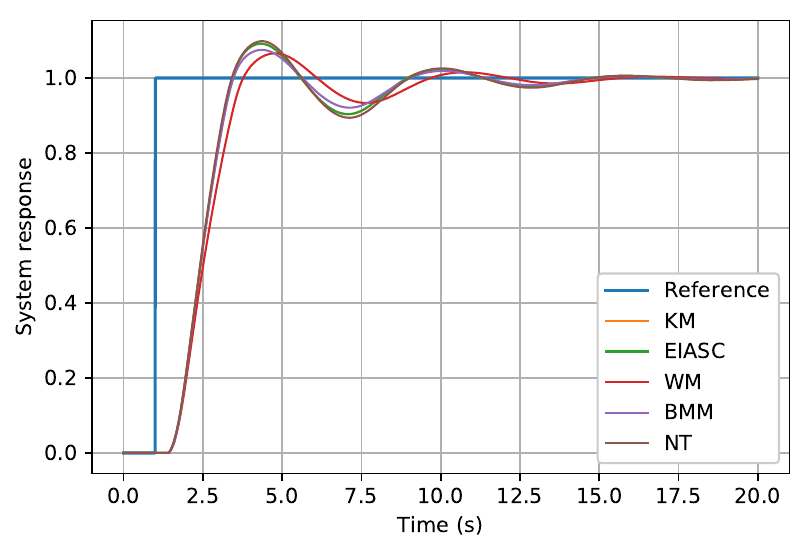}
\caption{Perturbed system 1.}
\label{FPID_Plant2}
\end{subfigure}%
\begin{subfigure}{0.32\linewidth}
\centering
\includegraphics[width=5cm]{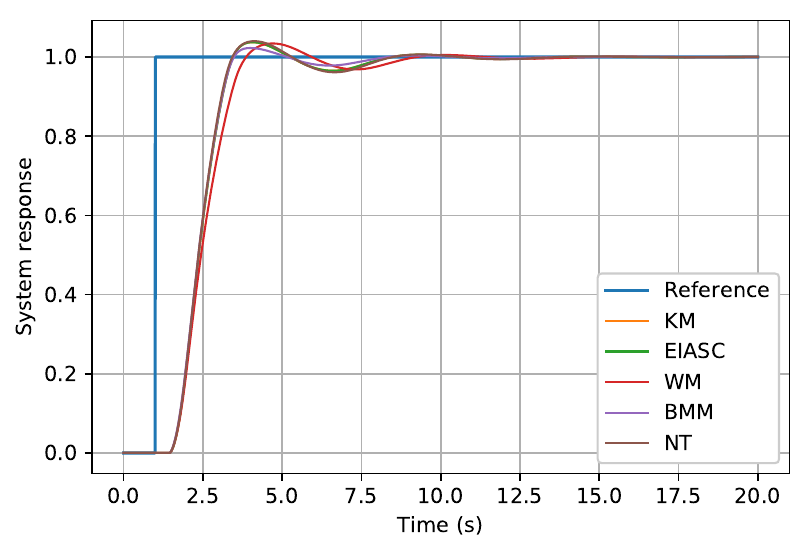}
\caption{Perturbed system 2.}
\label{FPID_Plant3}
\end{subfigure}\\[1ex]
\caption{The closed loop step response for main system, perturbed system-1, and perturbed system-2.}
\label{FPID_simuresult}
\end{figure}

In this section in order to show the generality of the proposed toolkit, it is used to model an IT2FPID controller. It must be noted that this case study is adopted from \cite{Taskin2015it2fls}. Let's consider the transfer function of the system under study as below:
\begin{equation}
G(s) = \frac{K}{Ts+1} e^{-Ls}
\end{equation}
And let the block diagram representation of the overall system be as illustrated in Figure. \ref{blockdiag}.
As it can be seen in Figure. \ref{blockdiag}, the inputs of the IT2FPID are the scaled verions of the error signal ($e$) and its derivative ($\dot{e}$). For each input there are three IT2FSs, Negative (N), Zero (Z), and Positive (P). The single output of the IT2FPID is used with scaling factors for achieving the proportional and integral terms coefficients. The output is demonstrated using five IT2FSs, Negative Big (NB), Negative Medium (NM), Zero (Z), Positive Medium (PM), and Positive Big (PB). All input and output sets are Gaussian ones with uncertain standard deviation. Also, the universe of discourse is defined as the interval $[-1,\ 1]$. Input and output sets are illustrated in Figures \ref{i_IT2FPID_Sets} and \ref{o_IT2FPID_Sets}, respectively.
The IT2FLS is constructed using 9 rules which are shown in Table. \ref{IT2FPID_ifthen}. 
\begin{table}
 \caption{Fuzzy If-Then rule base of IT2FPID.}
  \centering
  \begin{tabular}{cccc}
    \hline
    $\dot{e} / e$ &N &Z &P \\
    \hline
    N & NB & NM & Z \\
    Z & NM & Z & PM \\
    P & Z & NM & PB\\
    \hline
  \end{tabular}
  \label{IT2FPID_ifthen}
\end{table}
Parameters of the nominal system are set as $L=0.2$, $T=1$, and $K=1$. In order to study the effect of parameters uncertainty on the controller's performance, two more parameter settings are considered; $L=0.4$, $T=1.9$, $K=1.3$ (Perturbed System-1) and $L=0.45$, $T=1.3$, $K=1.1$ (Perturbed System-2). The scaling factors of the IT2FPID are set and fixed as $K_{a} = 0.25$, $K_{b} = 4.25$, $K_{e} = 0.8$, and $K_{d} = 0.5$. The performance of the designed control system is measured based on overshoot, settling time, and Integral Time Absolute Error (ITAE) values. The step response of the systems are illustrated in Figure. \ref{FPID_simuresult}, and the performance measures are reported in Table. \ref{performancetable}. As it can be seen, interval type 2 fuzzy logic system improves the robustness of the system response faced with system parameters perturbation. The simulations are done with multiple type reduction algorithms alongside with Centroid method. The results show that WM and BMM type reduction algorithms demonstrate better performance facing with control problems.

\begin{table}
 \caption{Performance measure of designed IT2FPID in contact with system.}
  \centering
  \begin{tabular}{ccccc}
    \hline
    System &TR &Settling time &Overshoot &ITAE \\
    \hline
    Nominal system 		&KM &5.1125 &2.2287\% &2.7967 \\
    					&EIASC &5.1225 &2.2302\% &2.7970 \\
    					&WM &5.3926 &2.0904\% &3.2899 \\
    					&BMM &5.0925 &2.1204\% &2.8455 \\
    					&NT &5.1325 &2.5284\% &2.7554 \\
    \hline
    Perturbed system-1 	&KM &13.1765 &9.3269\% &5.9002 \\
        				&EIASC &13.1765 &9.3291\% &5.8978 \\
    					&WM &9.0545 &6.8808\% &5.4139 \\
    					&BMM &10.2951 &7.6610\% &5.3664 \\
    					&NT &13.4267 &10.0322\% &6.2336 \\
    \hline
    Perturbed system-2 	&KM &7.5437 &3.7081\% &3.3893 \\
        				&EIASC &7.5437 &3.7139\% &3.3908 \\
    					&WM &8.1940 &3.4181\% &3.7504 \\
    					&BMM &6.8834 &2.2547\% &3.0258 \\
    					&NT &7.6738 &3.9807\% &3.4711 \\
    \hline
  \end{tabular}
  \label{performancetable}
\end{table}

\section{Conclusion}
In this paper a new Python toolkit for utilizing interval type 2 fuzzy logic systems is presented. This toolkit facilitates the use of IT2FLSs in practical scientific and engineering applications. How to use this library and its features are briefly described in this paper. Also, several examples are included for better understanding. Prediction of time series is one of the most common applications of fuzzy systems. PyIT2FLS used to predict the chaotic Mackey-Glass time series. Also a IT2FPID controller designed to control a time delay linear system. As the results show, the presented toolkit is able to be used in complex applications of the IT2FLSs. The development of the provided toolkit here is not over, but will continue to cover a wider variety of systems, methods and applications. Compared with other Python toolkits for type 2 fuzzy logic, PyIT2FLS supports both fuzzy sets and fuzzy systems and that is while type 2 fuzzy systems are not supported by other Python toolkits. It should be noted that the developers of the PyIT2FLS are ready to receive comments and suggestions about the toolkit and possible bug reports. The PyIT2FLS is openly accessible on the GitHub page: \url{https://github.com/Haghrah/PyIT2FLS}. 

\section*{Appendix: Python codes}
Simple python codes used in the paper for generating outputs are provided here. The codes for the two last complete examples are accessible in PyIT2FLS GitHub page.

\subsection*{Defining an IT2FS using PyIT2FLS}
\begin{python}
from pyit2fls import IT2FS, trapezoid_mf, tri_mf
from numpy import linspace

mySet = IT2FS(linspace(0., 1., 100), 
              trapezoid_mf, [0, 0.4, 0.6, 1., 1.], 
              tri_mf, [0.25, 0.5, 0.75, 0.6])
\end{python}
\subsection*{Plotting the meet and join of IT2FSs}
\begin{python}
from pyit2fls import IT2FS_Gaussian_UncertMean, IT2FS_plot, meet, \
                     join, min_t_norm, max_s_norm
from numpy import linspace

domain = linspace(0., 1., 100)

A = IT2FS_Gaussian_UncertMean(domain, [0., 0.1, 0.1])
B = IT2FS_Gaussian_UncertMean(domain, [0.33, 0.1, 0.1])
C = IT2FS_Gaussian_UncertMean(domain, [0.66, 0.1, 0.1])

IT2FS_plot(A, B, C, title="", legends=["Small","Medium","Large"], filename="multiSet")

AB = meet(domain, A, B, min_t_norm)
AB.plot(filename="meet")

BC = join(domain, B, C, max_s_norm)
BC.plot(filename="join")
\end{python}

\subsection*{Plotting the type reduced output of IT2FLS and calculating the crisp output}
\begin{python}
TR_plot(domain, TR["y1"], filename="y1_tr")
print(crisp(TR["y1"]))

TR_plot(domain, TR["y2"], filename="y2_tr")
print(crisp(TR["y2"]))
\end{python}

\subsection*{Full example of defining and using an IT2FLS}
\begin{python}
from pyit2fls import IT2FLS, IT2FS_Gaussian_UncertStd, IT2FS_plot, \
                     min_t_norm, max_s_norm, TR_plot
from numpy import linspace

domain = linspace(0., 1., 100)

Small = IT2FS_Gaussian_UncertStd(domain, [0, 0.15, 0.1])
Medium = IT2FS_Gaussian_UncertStd(domain, [0.5, 0.15, 0.1])
Large = IT2FS_Gaussian_UncertStd(domain, [1., 0.15, 0.1])
IT2FS_plot(Small, Medium, Large, legends=["Small", "Medium", "large"], filename="simp_ex_sets")

myIT2FLS = IT2FLS()
myIT2FLS.add_input_variable("x1")
myIT2FLS.add_input_variable("x2")
myIT2FLS.add_output_variable("y1")
myIT2FLS.add_output_variable("y2")

myIT2FLS.add_rule([("x1", Small), ("x2", Small)], [("y1", Small), ("y2", Large)])
myIT2FLS.add_rule([("x1", Medium), ("x2", Medium)], [("y1", Medium), ("y2", Small)])
myIT2FLS.add_rule([("x1", Large), ("x2", Large)], [("y1", Large), ("y2", Small)])

it2out, tr = myIT2FLS.evaluate({"x1":0.9, "x2":0.9}, min_t_norm, max_s_norm, 
                      domain, method="Centroid")

it2out["y1"].plot(filename="y1_out")
TR_plot(domain, tr["y1"], filename="y1_tr")

it2out["y2"].plot(filename="y2_out")
TR_plot(domain, tr["y2"], filename="y2_tr")
\end{python}

\bibliographystyle{unsrt}  
\bibliography{references}

\begin{thebibliography}{10}

\bibitem{tanaka2004fuzzy}
Kazuo Tanaka and Hua~O Wang.
\newblock {\em Fuzzy control systems design and analysis: a linear matrix
  inequality approach}, page~5.
\newblock John Wiley \& Sons, 2004.

\bibitem{wang1997course}
Li-Xin Wang and Li-Xin Wang.
\newblock {\em A course in fuzzy systems and control}, volume~2, pages 2--3.
\newblock Prentice Hall PTR Upper Saddle River, NJ, 1997.

\bibitem{zadeh1975concept}
Lotfi~Asker Zadeh.
\newblock The concept of a linguistic variable and its application to
  approximate reasoning—i.
\newblock {\em Information sciences}, 8(3):199--249, 1975.

\bibitem{mendel2014introduction}
Jerry Mendel, Hani Hagras, Woei-Wan Tan, William~W Melek, and Hao Ying.
\newblock {\em Introduction to type-2 fuzzy logic control: theory and
  applications}, page~43.
\newblock John Wiley \& Sons, 2014.

\bibitem{atacak2012type}
Ismail Atacak and Omer~Faruk Bay.
\newblock A type-2 fuzzy logic controller design for buck and boost dc--dc
  converters.
\newblock {\em Journal of intelligent manufacturing}, 23(4):1023--1034, 2012.

\bibitem{mendez2010modelling}
GM~Mendez, L~Leduc-Lezama, R~Colas, G~Murillo-Perez, J~Ramirez-Cuellar, and
  JJ~Lopez.
\newblock Modelling and control of coiling entry temperature using interval
  type-2 fuzzy logic systems.
\newblock {\em Ironmaking \& Steelmaking}, 37(2):126--134, 2010.

\bibitem{cheng2010inverse}
Li~Cheng-Dong, Yi~Jian-Qiang, Yu~Yi, and Zhao Dong-Bin.
\newblock Inverse control of cable-driven parallel mechanism using type-2 fuzzy
  neural network.
\newblock {\em Acta Automatica Sinica}, 36(3):459--464, 2010.

\bibitem{barkati2008application}
S~Barkati, EM~Berkouk, and MS~Boucherit.
\newblock Application of type-2 fuzzy logic controller to an induction motor
  drive with seven-level diode-clamped inverter and controlled infeed.
\newblock {\em Electrical Engineering}, 90(5):347--359, 2008.

\bibitem{singh2018adaptive}
Vikas Singh, Raghav Dev, Narendra~K Dhar, Pooja Agrawal, and Nishchal~K Verma.
\newblock Adaptive type-2 fuzzy approach for filtering salt and pepper noise in
  grayscale images.
\newblock {\em IEEE Transactions on Fuzzy Systems}, 26(5):3170--3176, 2018.

\bibitem{wagner2013juzzy}
Christian Wagner.
\newblock Juzzy-a java based toolkit for type-2 fuzzy logic.
\newblock In {\em 2013 IEEE Symposium on Advances in Type-2 Fuzzy Logic Systems
  (T2FUZZ)}, pages 45--52. IEEE, 2013.

\bibitem{castillo2013computational}
Oscar Castillo, Patricia Melin, and Juan~R Castro.
\newblock Computational intelligence software for interval type-2 fuzzy logic.
\newblock {\em Computer Applications in Engineering Education}, 21(4):737--747,
  2013.

\bibitem{Taskin2015it2fls}
Ahmet Taskin and Tufan Kumbasar.
\newblock An open source matlab/simulink toolbox for interval type-2 fuzzy
  logic systems.
\newblock In {\em 2015 IEEE Symposium Series on Computational Intelligence},
  pages 1561--1568. IEEE, 2015.

\bibitem{McCulloch2017}
Josie McCulloch.
\newblock Fuzzycreator: A python-based toolkit for automatically generating and
  analysing data-driven fuzzy sets.
\newblock In {\em 2017 IEEE International Conference on Fuzzy Systems
  (FUZZ-IEEE)}, pages 1--6. IEEE, 2017.

\bibitem{oliphant2006guide}
Travis~E Oliphant.
\newblock {\em A guide to NumPy}, volume~1.
\newblock Trelgol Publishing USA, 2006.

\bibitem{hunter2007matplotlib}
John~D Hunter.
\newblock Matplotlib: A 2d graphics environment.
\newblock {\em Computing in science \& engineering}, 9(3):90, 2007.

\bibitem{karnik1999type}
Nilesh~Naval Karnik, Jerry~M Mendel, and Qilian Liang.
\newblock Type-2 fuzzy logic systems.
\newblock {\em IEEE transactions on Fuzzy Systems}, 7(6):643--658, 1999.

\bibitem{duran2008improved}
K~Duran, H~Bernal, and M~Melgarejo.
\newblock Improved iterative algorithm for computing the generalized centroid
  of an interval type-2 fuzzy set.
\newblock In {\em NAFIPS 2008-2008 Annual Meeting of the North American Fuzzy
  Information Processing Society}, pages 1--5. IEEE, 2008.

\bibitem{wu2008enhanced}
Dongrui Wu and Jerry~M Mendel.
\newblock Enhanced karnik--mendel algorithms.
\newblock {\em IEEE Transactions on Fuzzy Systems}, 17(4):923--934, 2008.

\bibitem{wu2011comparison}
Dongrui Wu and Maowen Nie.
\newblock Comparison and practical implementation of type-reduction algorithms
  for type-2 fuzzy sets and systems.
\newblock In {\em 2011 IEEE International Conference on Fuzzy Systems
  (FUZZ-IEEE 2011)}, pages 2131--2138. IEEE, 2011.

\bibitem{wu2002uncertainty}
Hongwei Wu and Jerry~M Mendel.
\newblock Uncertainty bounds and their use in the design of interval type-2
  fuzzy logic systems.
\newblock {\em IEEE Transactions on fuzzy systems}, 10(5):622--639, 2002.

\bibitem{begian2008stability}
Mohammad~Biglar Begian, William~W Melek, and Jerry~M Mendel.
\newblock Stability analysis of type-2 fuzzy systems.
\newblock In {\em 2008 IEEE International Conference on Fuzzy Systems (IEEE
  World Congress on Computational Intelligence)}, pages 947--953. IEEE, 2008.

\bibitem{li2010stability}
Chengdong Li, Jianqiang Yi, and Tiechao Wang.
\newblock Stability analysis of sirms based type-2 fuzzy logic control systems.
\newblock In {\em International Conference on Fuzzy Systems}, pages 1--7. IEEE,
  2010.

\bibitem{nie2008towards}
Maowen Nie and Woei~Wan Tan.
\newblock Towards an efficient type-reduction method for interval type-2 fuzzy
  logic systems.
\newblock In {\em 2008 IEEE International Conference on Fuzzy Systems (IEEE
  World Congress on Computational Intelligence)}, pages 1425--1432. IEEE, 2008.

\bibitem{glass2010mackey}
Leon Glass and Michael Mackey.
\newblock Mackey-glass equation.
\newblock {\em Scholarpedia}, 5(3):6908, 2010.

\bibitem{roussel2018mackey}
Marc~R Roussel.
\newblock The mackey-glass models, 40 years later.
\newblock {\em Biomath Communications}, 5(2):140--158, 2018.

\bibitem{zhongda2017prediction}
Tian Zhongda, Li~Shujiang, Wang Yanhong, and Sha Yi.
\newblock A prediction method based on wavelet transform and multiple models
  fusion for chaotic time series.
\newblock {\em Chaos, Solitons \& Fractals}, 98:158--172, 2017.

\bibitem{heydari2016chaotic}
Gholamali Heydari, MohammadAli Vali, and Ali~Akbar Gharaveisi.
\newblock Chaotic time series prediction via artificial neural square fuzzy
  inference system.
\newblock {\em Expert Systems with Applications}, 55:461--468, 2016.

\bibitem{shahverdiev2006chaos}
EM~Shahverdiev, RA~Nuriev, RH~Hashimov, and KA~Shore.
\newblock Chaos synchronization between the mackey--glass systems with multiple
  time delays.
\newblock {\em Chaos, Solitons \& Fractals}, 29(4):854--861, 2006.

\bibitem{shi2017synchronization}
Hong-jun Shi, Lian-ying Miao, and Yong-zheng Sun.
\newblock Synchronization of time-delayed systems with discontinuous coupling.
\newblock {\em Kybernetika}, 53(5):765--779, 2017.

\end{thebibliography}

\end{document}